\documentclass[useAMS,usenatbib]{mn2e}

\usepackage{amssymb,amsmath,upgreek,bm,color}

\usepackage{epsfig,graphicx,times,subfigure}
\usepackage[colorlinks=true, citecolor=cyan,
 linkcolor=cyan, menucolor=blue, urlcolor=blue,
 linkbordercolor={0 0 1}, frenchlinks=True, breaklinks]{hyperref}

\newcommand{\aaps}{A\&AS}
\newcommand{\aap}{A\&A}
\newcommand{\aj}{AJ}
\newcommand{\pasp}{PASP}
\newcommand{\apj}{ApJ}

\newcommand{\mnras}{MNRAS}
\newcommand{\araa}{ARA\&A}
\newcommand{\apjs}{ApJS}
\newcommand{\nat}{Nature}

\newcommand{\oiii}{\hbox{[O\,{\sc iii}]}}
\newcommand{\ha}{\hbox{H$\alpha$}}
\newcommand{\hb}{\hbox{H$\beta$}}
\newcommand{\oii}{\hbox{[O\,{\sc ii}]}}
\newcommand{\sii}{\hbox{[S\,{\sc ii}]}}
\newcommand{\nii}{\hbox{[N\,{\sc ii}]}}

\usepackage{gensymb}

\title[An Edge-on Galaxy with X-shaped Bi-cone]
{Discovery of an Edge-on Galaxy with X-shaped Bi-cone --- SDSS J171359.00+333625.5}

\author[Bao, M. et al.]{
\parbox[t]{\textwidth}{\raggedright
Min Bao$^{1}$,
Yan-mei Chen$^{2}$\thanks{E-mail: chenym@nju.edu.cn},
Qi-rong Yuan$^{1}$\thanks{E-mail: yuanqirong@njnu.edu.cn},
Yong Shi$^{2}$,
Dmitry Bizyaev$^{3,4}$,
Xiao-ling Yu$^{2}$,
Qiu-sheng Gu$^{2}$,
Ying Yu$^{2}$
}\\
\vspace*{6pt}\\
$^{1}$Department of Physics and Institute of Theoretical Physics, Nanjing Normal University, Nanjing 210023, China\\
$^2$Department of Astronomy, Nanjing University, Nanjing 210093, China\\ 
    Key Laboratory of Modern Astronomy and  Astrophysics (Nanjing University), Ministry of Education, Nanjing 210093, China\\
    Collaborative Innovation Center of Modern Astronomy and Space Exploration, Nanjing 210093, China\\
$^3$Apache Point Observatory and New Mexico  State University, P.O.  Box 59, Sunspot, NM,  88349-0059,  USA\\
$^4$Sternberg Astronomical Institute, Moscow State University, Moscow, Russia\\
	}

\begin{document}

\pagerange{\pageref{firstpage}--\pageref{lastpage}} \pubyear{}

\maketitle
\newpage
\label{firstpage}
\newpage
\pagebreak

\begin{abstract}
Using the integral field unit (IFU) data from Mapping Nearby Galaxies at Apache Point Observatory (MaNGA) survey, we study the kinematics of gas and stellar components in an edge-on Seyfert 2 galaxy, SDSS J171359.00+333625.5, with X-shaped bi-conical outflows.  The gas and stars therein are found to be counter-rotating, indicating that the collision between the inner and external gas might be an effective way to dissipate the angular momentum, which leads to remarkable gas accretion into the galaxy center.  Large {\oiii}$\lambda$5007 equivalent width and AGN-like line ratio in the large bi-conical region suggest that the gas is ionized by the central AGN. The gas velocity in the bi-cone region shows that ionized gas is receding relative to the galaxy center, which could be the joint effect of inflows, outflows and disk rotation.  We are probably witnessing the case where a great amount of gas in the disk is being efficiently accreted into the central black hole, and the AGN-driven galactic winds are blown out along the bi-cone. The kinematics of oxygen, including rotation velocity and velocity dispersion, is different from other elements, like hydrogen, nitrogen and sulfur. The rotation velocity estimated from oxygen is slower than from other elements. The velocity dispersion of other elements follows galactic gravitational potential, while the velocity dispersion of oxygen stays roughly constant along the galactic major-axis. The further advanced observations, e.g. of cold gas or with an IFU of higher spatial resolution, are required to better understand this object.
\end{abstract}


\section{Introduction}
\label{Introduction}
The regulation of star formation in galaxies and building-up of their stellar mass are hot topics and major unresolved issues in the research of the galaxy formation and evolution. Galactic scale winds in cooperation with the gas feeding and the star formation efficiency are believed to be processes important for understanding of how galaxies evolve. 

About twenty years ago, two seminal papers \citep{1998AJ....115.2285M, 1998A&A...331L...1S} proposed the idea that the energy released from an active galactic nucleus (AGN) can heat up and blow out the cold gas as well as effectively quench the formation of new stars, in the case of efficient coupling with the interstellar medium in its host galaxy. Over the last two decades, lots of efforts have been made in searching for the AGN feedback evidence \citep{2017ApJ...834...30F, 2019ApJ...870...37D}, including quantifying the importance of feedback process \citep{2007ApJ...656..699D, 2017ApJ...844...37D, 2018MNRAS.473.4077P} in galaxy evolution. Nevertheless, consensus has not been reached on the effect of AGN feedback, which is associated with both suppressing and triggering the star formation. In theory, massive and fast outflows can suppress the star formation in the host galaxy by removing and heating the ISM. Near-IR IFU observations of $z \sim$ 1 $-$ 3 quasars have revealed a spatial anti-correlation between the location of the fast outflowing gas and the star formation in the host galaxy\citep{2012A&A...537L...8C, 2015ApJ...799...82C, 2016A&A...591A..28C}. However, only a small fraction of the outflowing gas may escape the host halo, while a large fraction may fall back onto the galaxy at later times\citep{2014A&A...568A..14A}. The energy and angular momentum carried by this part of gas may be injected into the halo, which can heat the gas and prevent it from cooling, as well as halt the accretion of the fresh gas. Moreover, not all forms of feedback suppress the star formation. Fast outflows can also induce the star formation in the galactic disk \citep{2013ApJ...772..112S} or in the galactic winds \citep{2012MNRAS.427.2998I} through the compression of molecular clouds.

The wealth of observational data from the most advanced facilities point towards the complexity of physical conditions in galactic winds. First of all, most of the AGNs in the local universe are hosted by the spiral galaxies with non-ignorable star formation. It is unclear whether AGN or star formation activity primarily drive the winds \citep{2005ARA&A..43..769V}. Secondly, different physical mechanisms, i.e. the radiative energy and/or the mechanical effects \citep{2008MNRAS.383..119D}, can also drive galactic outflows. It is often unclear which mechanism is dominant for the winds, and if we are able to disentangle different mechanisms. Identifying the primary driver and physical mechanism of galactic-scale gaseous winds requires progress in the quality of available data, which should include multi-wavelength information as well as spatial resolved spectra.

In the cases of spatially resolved spectroscopic observations, the geometry of the outflows may be helpful for assessing the dominant driving process. For example, \cite{2015ApJ...806...84L} showed that the base of the outflow in a low-luminosity Seyfert galaxy is coincident with the unresolved nucleus, supporting the hypothesis that AGN is indeed the predominant ionizing source of the outflowing gas. \cite{2010ApJ...721..505R} found that gas in NGC 839 is concentrated in a bi-conical polar funnel, which can be interpreted as shock-excited superwind. An integral field spectroscopic (IFS) study of two nearby luminous infrared galaxies (LIRGs), IC 1623 and NGC 3256, exhibited the evidence of widespread shock excitation induced by ongoing merger activity \citep{2011ApJ...734...87R}. Optical IFS observations of the Mice, a major merger between two massive gas-rich spirals NGC 4676A and B, uncovered that both galaxies show bi-cones of the high ionized gas extending along their minor-axis \citep{2014A&A...567A.132W}. As in the cases above, the AGN, shock or merger activity can work as the driving mechanism for outflows. The common characteristic in outflow is the presence of ionized gas in the bi-cones.

Using long-slit spectra from the Hubble Space Telescope (HST), \cite{2005AJ....130..945D} presented detailed information on a face-on galaxy with ionized bi-cone, NGC4151, and proposed a model with bi-conical NLR (see Figure 12 in \cite{2005AJ....130..945D}). The inclination of the galaxy and bi-cone are 20$^{\circ}$ and 45$^{\circ}$, respectively, thus the intersection angle is only 25$^{\circ}$. \cite{2010AJ....140..577F} and \cite{2010AJ....139..871C} developed geometric models of the NLRs and the inner disks of MRK 573 and MRK 3. Both models uncovered the possibility that the inner disk can spatially traverse the cones in both sides. It can be inferred that a certain number of galaxies with bi-conical NLRs are different from the classical model of M82 where the bi-cone is exactly perpendicular to the disk. 

More than 50 years ago, \citet{1963ApJ...137.1005L} discovered an evidence of an explosion at the center of M82. Prior to this, few works discussed the galactic winds. The progress was restricted because of the lack of comprehensive data over the full electromagnetic spectra at comparable sensitivity and spatial resolution. Observations of the entire electromagnetic spectrum has become possible over the last fifty years. Spatially resolved, spectral maps provided by the IFS enable us to study galactic winds in detail. Recently, based on the survey of MaNGA, we catch an edge-on galaxy, SDSS J171359.00+333625.5, with AGN-driven winds. The discovery of galactic winds with small fibre bundle is great news for the multi-bundle instruments like SAMI \citep{2012ApJ...761..169F}, MANGA as well as the next big version Hector. We discover clear bi-conical outflows in this case, as well as detailed gas and stellar kinematics, which offers an excellent opportunity to study fueling and feedback processes. The sample selection and methodology of data processing is presented in Section \ref{sec:data}. We show the kinematics of gas and stars, as well as the ionization status in bi-cone in Section \ref{sec:properties}. Our understandings on these results and two unusual properties of this galaxy are discussed in Section \ref{sec:discussion}. Finally, a summary is given in Section \ref{sec:summary}. Throughout this paper, we adopt a set of cosmological parameters as follows: $H_0=70\,{\rm km~s}^{-1}\,{\rm Mpc}^{-1}$, $\Omega_m=0.30$, $\Omega_{\Lambda}=0.70$.

\section{Data Analysis}
\label{sec:data}

MaNGA is one of three core programs in the fourth-generation Sloan Digital Sky Survey (SDSS-IV) \citep{2015ApJ...798....7B, 2016AJ....152...83L}. MaNGA employs the Baryon Oscillation Spectroscopic Survey (BOSS) spectrographs \citep{2013AJ....146...32S} on the 2.5m Sloan Foundation Telescope \citep{2006AJ....131.2332G}. This survey aims to conduct IFU observation for a representative sample, which consists of about 10,000 nearby galaxies with stellar mass $\log (M_{\ast}/M_{\odot}$) $\geq$ 9 and redshift $0.01 < z < 0.15$ \citep{2017AJ....154...28B}. Two dual-channel BOSS spectrographs \citep{2013AJ....146...32S} provide simultaneous wavelength coverage from 3600 to 10,000$\rm \AA$. The spectral resolution is $\sim$2000, allowing measurements of all strong emission line species from {\oii}$\lambda$3727 to {\sii}$\lambda$6731. The MaNGA sample and data products used here were drawn from the SDSS Data Release 15 (DR15) \citep{2019ApJS..240...23A}, which includes $\sim$4633 galaxies observed through July 2016 (the first three years of the survey). 

Using the demarcation\\
$\log$({\oiii}$\lambda5007/H\rm \beta$)$=$0.61$/$($\log$({\nii}$\lambda6583/\rm H\alpha$)$-$0.47)$+$1.19 \citep{2001ApJ...556..121K}, we firstly select the AGN galaxies from the MaNGA sample. After that we use the criterion\\
$\log$({\oiii}$\lambda5007/\rm H\beta$)$=$1.05$\log$({\nii}$\lambda6583/\rm H\alpha$)$+$0.45 \citep{2007MNRAS.382.1415S}, and further select 113 Seyfert galaxies. We then inspect the {\oiii}$\lambda$5007 and H$\alpha$ equivalent width (EQW) maps of these Seyfert galaxies by eyes and find 13 of them having bi-conical features in {\oiii}$\lambda$5007 EQW maps. Among these 13 galaxies, SDSS J171359.00+333625.5 is a Seyfert 2 galaxy with the spectroscopic redshift of $z \sim$ 0.039, and it has the most prominent bi-cone feature as well as interesting gas and stellar kinematics. In this paper we will focus on this object and explore its unusual properties. 

The MaNGA data reduction pipeline (DRP) \citep{2016AJ....152...83L} provides spectral datacubes. We reanalyze the DRP spectra of this object using the principal component analysis (PCA) method as described in \cite{2016MNRAS.463..913J}. We directly use the principal components (PCs) and library of model spectra given by \cite{2012MNRAS.421..314C} to get the velocity field of stars. We shift the best-fitting model from $-$1000 km s$^{-1}$ to 1000 km s$^{-1}$ by a step size of 2 km s$^{-1}$. For each step, we calculate the reduced $\chi^{2}$ between the best-fitting model and the observed spectrum. The velocity of stars at a certain spaxel is determined by the fit with the lowest $\chi^{2}$ value. After that, we model the stellar continuum of each spectrum using the BC03 \citep{2003MNRAS.344.1000B} stellar population synthesis models and separate the stellar continuum and absorption lines from the nebular emission lines. The best-fitting continuum model is then subtracted from the observed spectrum to give the pure emission line spectrum. Each emission line is then fitted with one Gaussian component using MPFIT \citep{2009ASPC..411..251M}. We include {\oii}$\lambda$3727, {\hb}, {\oiii}$\lambda\lambda$4959,5007, {\ha}, {\nii}$\lambda\lambda$6548,6584, {\sii}$\lambda\lambda$6717,6731 emission lines in our line fitting. The center and width of these lines are not tied with each other. We limit the shift of the line center to be in the range of [$-$300, 300] km s$^{-1}$ after redshift correction. The line widths are limited by the range of [0, 500] km s$^{-1}$.

\section{Properties of SDSS J171359.00+333625.5}
\label{sec:properties}

\subsection{Kinematics}
\subsubsection{\rm Counter-rotating Gas and Stellar Components}
\label{Subsec:counter-rotator}
\begin{figure*}
 \resizebox{1.\textwidth}{!}{\includegraphics{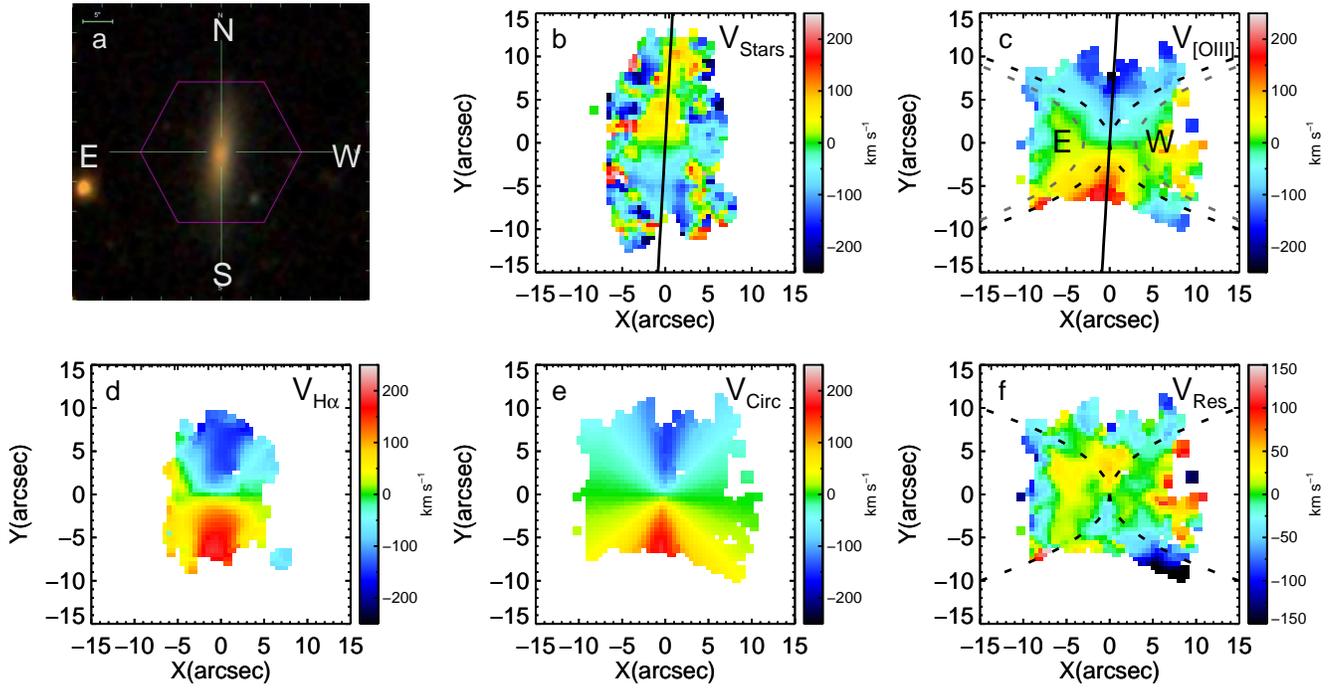}}
\caption{a: SDSS g, r, i color image of the galaxy. The purple hexagon shows the MaNGA bundle allocation. b: The stellar velocity field for spaxels with spectral S/N large than 2 per pixel. The black solid line represents the major-axis of stellar rotation. c: The {\oiii}$\lambda$5007 velocity field for spaxels with {\oiii}$\lambda$5007 S/N larger than 3. The black solid line represents the major-axis of gas rotation. The grey hyperbola shows the edge of the kinematic bi-cone, the black hyperbola shows the edge of photo-ionized bi-cone. d: The {\ha} velocity field for spaxels with {\ha} emission line S/N larger than 3. e: The circular disk model. f: Residual (V$_{\rm [OIII]}$ - V$_{\rm Circ}$) after subtracting the disk modal from the {\oiii}$\lambda$5007 velocity field. The black hyperbola outlines the edge of ionized bi-cone.}
\label{fig1}
\end{figure*}

In Figure \ref{fig1}, we show the gas and stellar kinematics of SDSS J171359.00+333625.5. Figure \ref{fig1}a is the SDSS $g,r,i$ color image. Figure \ref{fig1}b shows the stellar velocity field for the spaxels with spectral signal-to-noise ratio (S/N) larger than 2 per pixel. Figure \ref{fig1}c shows the {\oiii}$\lambda$5007 velocity field for spaxels with {\oiii}$\lambda$5007 S/N larger than 3. Figure \ref{fig1}d shows the {\ha} velocity field for spaxels with H$\alpha$ emission line S/N larger than 3. 
The position angles of gas and stellar velocity fields are measured using the IDL package \textsc{\small KINEMETRY} \citep{2006MNRAS.366..787K}. The position angle (PA) is defined as the counter-clockwise angle between north and a line that bisects the velocity field of gas or stars on the receding side. The black solid lines in Figure \ref{fig1}b and \ref{fig1}c mark the PAs of gas and stellar velocity fields. The kinematic misalignment between gas and stars is defined as $\Delta$PA = $|$PA$_{\rm gas}$ - PA$_{\ast}|$, where PA$_{\rm gas}$ is the PA of ionized gas and PA$_{\ast}$ is the PA of stars. It turns out that SDSS J171359.00+333625.5 is a counter-rotator with $\Delta$PA $\sim$ 179.8$^{\circ}$. 
Comparing the rotation in the {\oiii}$\lambda$5007 and H$\alpha$ velocity fields, 
we find that the H$\alpha$ tends to trace the gaseous disk, while the {\oiii}$\lambda$5007 extends to the bi-cone region. We then build a pure circular disk rotation field expressed as 
\begin{equation}
V(R,\psi) = V_{0} + V_{c}(R)\sin i\cos \psi,
\label{eqvr}
\end{equation}
where $R$ is the radius of a circular ring in the galactic plane, $V_{0}$ is the systemic velocity, $V_{c}$ is the circular velocity at radius $R$ and $\psi$ is the azimuthal angle measured from the major-axis in the galactic plane \citep{2006MNRAS.366..787K} with a coverage of [-$\pi/2$, $\pi/2$]. $V_{0}$ approaches zero after the redshift correction, and $V_{c}(R)$ is defined as the H$\alpha$ rotation velocity along the major-axis. Figure \ref{fig1}e shows the circular disk model. We subtract the disk model from {\oiii}$\lambda$5007 velocity field. The residuals are shown in Figure \ref{fig1}f. 

The inclination angle ($i$) is calculated using the algorithm given by \cite{1926ApJ....64..321H}, in which the observed axis ratio is compared to the intrinsic value $q_{0}$. We use \textsc{\small KINEMETRY} to fit the galaxy axis ratio ($q$), and the result is $q = 0.28$. 
The relation between the inclination angle and axial ratio is given by
\begin{equation}
\cos^{2}i = \frac{q^{2} - q_{0}^{2}}{1 - q_{0}^{2}}
\label{eqvr}
\end{equation}
\citep{1926ApJ....64..321H}, where $q_{0}$ = 0.2. 
By following this equation we estimate that SDSS J171359.00+333625.5 is a nearly edge-on galaxy with $i = 78^{\circ}$.

\cite{2016NatCo...713269C} identified nine blue star forming counter-rotators from MaNGA, whose gas and stars have $\Delta$PA $>$ 150$^{\circ}$. The central regions of blue counter-rotators have younger stellar populations and more intense, ongoing star formation than their outskirts \citep{2019ApJ...882..145B}. All these phenomena suggest a picture in which the progenitor accretes the counter-rotating gas from the cosmic web or gas-rich dwarf galaxies, 
followed by redistribution of the angular momentum through the gas-gas collisions between the external and pre-existing gas. These counter-rotators largely accelerate gas inflows, leading to the fast centrally-concentrated star formation. As the gas flows into the central regions and feeds the central black hole, we expect to see activation of the AGN. This is the case of SDSS J171359.00+333625.5, a Seyfert 2 galaxy with bolometric AGN luminosity of 1.19$\times$10$^{44}$ erg s$^{-1}$.

\subsubsection{\rm Different {\oiii}$\lambda$5007 and H$\alpha$ Kinematics}
\label{Different [OIII] and Ha kinematics}
\begin{figure*}
 \resizebox{0.7\textwidth}{!}{\includegraphics{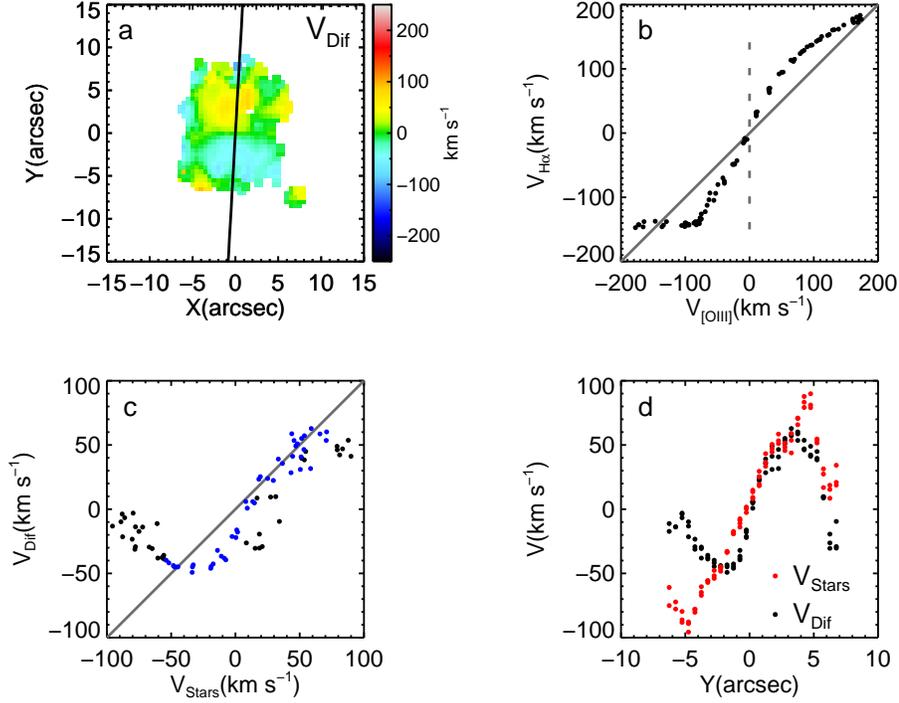}}
 \caption{a: The difference (V$_{\rm [OIII]}$ - V$_{\rm H\alpha}$) between the {\oiii}$\lambda$5007 and {\ha} velocity. The black solid line represents the major-axis. b: Relation between V$_{\rm [OIII]}$ and V$_{\rm H\alpha}$ along major-axis. The grey solid line shows the one-to-one correlation, and the grey dashed line marks the region where V$_{\rm [OIII]}$ equals to zero. c: Relation between the V$_{\rm Stars}$ and (V$_{\rm [OIII]}$ - V$_{\rm H\alpha}$) along major-axis. The grey solid line shows the one-to-one correlation. The blue dots represent spaxels with Y-position in the range of -3$\sim$4 arcsec, the black dots represent spaxels with Y-position over this range. d: The gas velocity as a function of the Y-position along major-axis. The red dots trace the velocity of stars, the black dots represent the residual (V$_{\rm [OIII]}$ - V$_{\rm H\alpha}$).}
 \label{fig2}
\end{figure*}

Figure \ref{fig1}c and \ref{fig1}d show an evidence that H$\alpha$ rotates in the same direction but faster than {\oiii}$\lambda$5007. To quantify the difference on rotation velocity between the {\oiii}$\lambda$5007 and H$\alpha$, we show the residual velocity (V$_{\rm [OIII]}$ $-$ V$_{\rm H\alpha}$) in Figure \ref{fig2}a. It is clear that this residual map is not a random fluctuation, but more resembles a regular rotation.
The rotation evidence can also be found in Figure \ref{fig1}f (at least along the major-axis), which shows the
residuals between {\oiii}$\lambda$5007 rotation and the circular disk model. The first impression is that the pattern of
residuals is consistent with kinematics of stars. To minimize the influence of gas kinematics in the bi-conical outflows, we focus on the spaxels located along the major axis. Figure \ref{fig2}b shows the relation between V$_{\rm [OIII]}$ and V$_{\rm H\alpha}$ along the major-axis, and it is obvious from this panel that the velocity of H$\alpha$ is higher than {\oiii}$\lambda$5007 at the same spaxels. We show the correlation between V$_{\rm Stars}$ and (V$_{\rm [OIII]}$ $-$ V$_{\rm H\alpha}$) along the major-axis in Figure \ref{fig2}c. The blue dots (with Y-position in the range of $-$3 $\sim$ 4 arcsec) roughly follow the one-to-one correlation. Over this range outlined with the black dots, the difference between {\oiii}$\lambda$5007 and H$\alpha$ decreases as the radius increases. In Figure \ref{fig2}d, we show (V$_{\rm [OIII]}$ $-$ V$_{\rm H\alpha}$) (black dots) and V$_{\rm Stars}$ (red dots) as a function of Y-position for clearer exhibition. In Figure \ref{fig2}d, we can see that the stellar velocity does not change smoothly with radii and it has a turnover at $|\rm Y| \sim 5$, which is roughly consistent with the turn-over position of
the difference (V$_{\rm [OIII]}$ $-$ V$_{\rm H\alpha}$).

\begin{figure*}
 \resizebox{1.\textwidth}{!}{\includegraphics{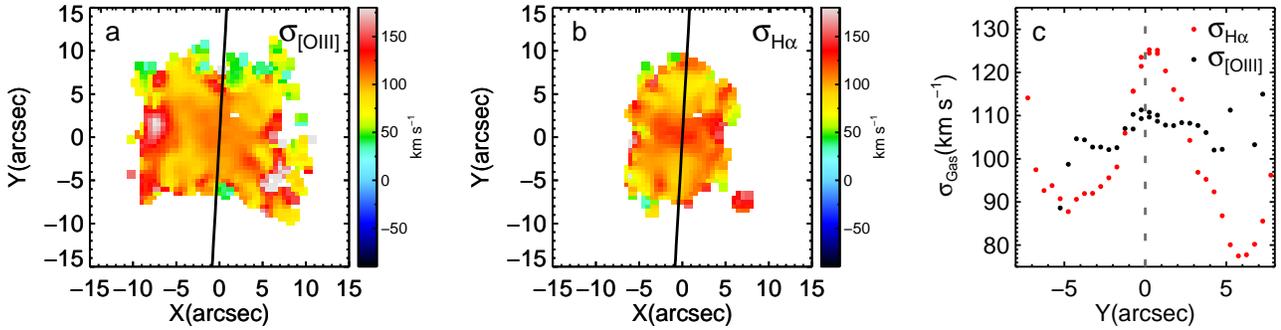}}
 \caption{a b: The velocity dispersion of the {\oiii}$\lambda$5007 and {\ha}. The black solid lines mark the major-axis. c: The gas velocity dispersion as a function of the Y-position along major-axis. The grey dashed line marks the regions where the Y-position equals zero. The red dots trace the velocity dispersion of {\ha}, while the black dots designate the velocity dispersion of the {\oiii}$\lambda$5007.}
 \label{fig3}
\end{figure*}

In Figure \ref{fig3}a and \ref{fig3}b, we show the velocity dispersion 
fields of {\oiii}$\lambda$5007 and H$\alpha$, respectively. We find that H$\alpha$ velocity dispersion peaks at the galaxy center while {\oiii}$\lambda$5007 peaks at the bi-cone region. Figure \ref{fig3}c shows the gas velocity dispersion as a function of the Y-position along major-axis. Again, we find different behaviors of {\oiii}$\lambda$5007 and H$\alpha$, that H$\alpha$ velocity dispersion peaks at the center and decreases as radius increases, while {\oiii}$\lambda$5007 velocity dispersion keeps roughly being constant of 100 km s$^{-1}$ along major-axis. 
We understand that the H$\alpha$ traces the galaxy potential well (with a bulge in the center and a disk in the outskirts), but the behavior of {\oiii}$\lambda$5007 is really wired in both velocity and velocity dispersion.

\begin{figure*}
 \resizebox{1.\textwidth}{!}{\includegraphics{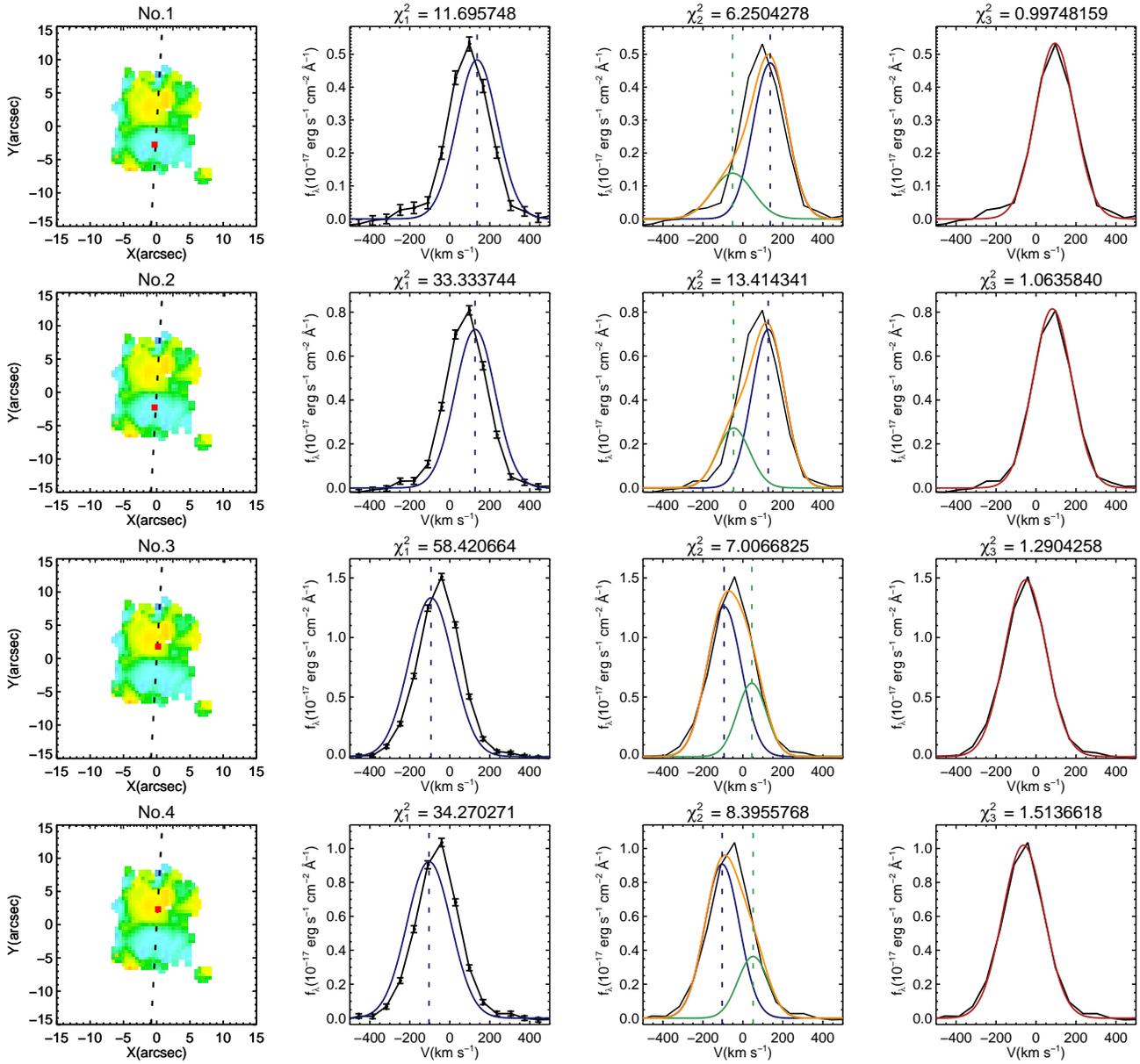}}
 \caption{The first column: The difference (V$_{\rm [OIII]}$ - V$_{\rm H\alpha}$) between the {\oiii}$\lambda$5007 and {\ha} velocity. The black dashed lines represent the major-axis. The red squares mark the pixels from which the spectra are subtracted. The second column: The {\oiii}$\lambda$5007 emission lines with error bar in each data points (black) and Gaussian fit (blue), which shares the same rotation velocity with {\ha}. The vertical blue dashed line corresponds to the velocity at the {\ha} line center. The third column: The black and blue profiles are same as in the second column. The Gaussian fit, which shares same rotation velocity with stars, is shown in green color. The orange profiles demonstrate the superposition of two Gaussian fits. The vertical blue and green dashed lines represent the velocity of {\ha} and stars, respectively. The last column: The black profiles are the same as in the second column. The red profiles show the single Gaussian fit with the shifted line center. The shift is limited by the range of [$-$300, 300] km s$^{-1}$.}
 \label{fig4}
\end{figure*}

To verify our conclusions, we check raw spectra from each MaNGA fiber. The result of our verification ensures that interpolation had negligible effect on both centers and widths of emission lines, so the difference between {\oiii}$\lambda$5007 and {\ha} is credible. Besides, we fit {\oiii}$\lambda$5007 in three schemes and the fitting results extracted from four pixels are plotted in Figure \ref{fig4} with the reduced chi-squared values labelled in the top of each panel. The specific positions of the four pixels are marked by red squares in the first column of Figure \ref{fig4}. In the second column, we fit a Gaussian (blue) to {\oiii}$\lambda$5007 and limit the velocity of line center same as {\ha}. Based on these four panels we find that the difference between {\oiii}$\lambda$5007 and {\ha} in the rotation velocity is apparent, and the chi-squared values ($\chi_{1}^{2}$) are much bigger compared to other two schemes ($\chi_{2}^{2}$ \& $\chi_{3}^{2}$). In the third column, we fit {\oiii}$\lambda$5007 with double Gaussians and limit the two lines' centers to share
the same rotation velocity with {\ha} (blue) and stars (green) respectively. If superposed profiles (orange) optimally depict the data, then the difference between {\oiii}$\lambda$5007 and {\ha} could be the result of collision between original and accreted clumps of gas.
The last column shows a Gaussian fit to the {\oiii}$\lambda$5007 line (red), 
with the only requirement that the line center displacement to range between $-$300
and 300 km s$^{-1}$. Comparing $\chi_{2}^{2}$ and $\chi_{3}^{2}$, we can tell that the best scheme is untied single Gaussian fit (in the last column), which means that the difference between {\oiii}$\lambda$5007 and {\ha} is caused by other kinematic processes instead of the gas collision.

We further check the behavior of {\oii}$\lambda\lambda$3726,3729, {\hb}, {\oiii}$\lambda$4959, {\nii}$\lambda\lambda$6548,6584 and {\sii}$\lambda\lambda$6716,6731, and find that the {\oii}$\lambda\lambda$3726,3729 and {\oiii}$\lambda$4959 have similar velocity and velocity dispersion to {\oiii}$\lambda$5007. At the same time, the {\hb}, {\nii}$\lambda\lambda$6548,6584 and {\sii}$\lambda\lambda$6716,6731 closely follow H$\alpha$. In summary, the kinematics of oxygen, including rotation velocity and velocity dispersion, is totally different from the other elements like hydrogen, nitrogen and sulphur. On the one hand, oxygen rotates slower than the other elements, although they share the same rotation direction; on the other hand, hydrogen, nitrogen and sulphur follow the galaxy potential well with a central bulge plus an outer disk while oxygen
does not.

\subsection{The Bi-conical, Narrow-line Region}
\label{subsec:bi-cone}
\begin{figure*}
 \resizebox{0.7\textwidth}{!}{\includegraphics{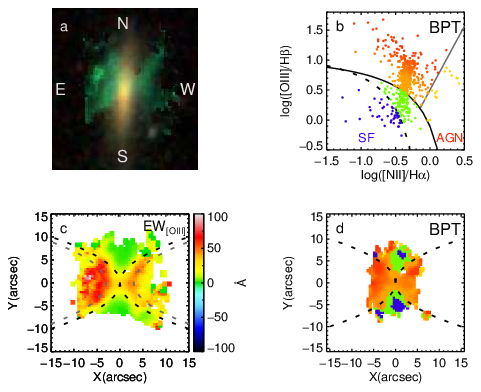}}
 \caption{a: The equivalent width map of {\oiii}$\lambda$5007 is plotted over the SDSS g, r, i color composite image. b: The BPT diagram for the spaxels where all S/Ns in the {\hb}, {\oiii}$\lambda$5007, {\ha} and {\nii}$\lambda$6583 are larger than 2. The black solid curve shows the demarcation between starburst galaxies and AGNs defined by \citet{2001ApJ...556..121K}, the black dashed curve shows the demarcation proposed by \citet{2003MNRAS.346.1055K}, while the grey solid line shows the separation between the shocks and AGNs defined by \citet{2010ApJ...711..818S}. The blue pixels designate the star forming region, the green pixels mark the composite region with contribution from both AGNs/shocks and star formation, the yellow pixels represent the shock region, while the orange pixels highlight the AGN region. c: The equivalent width of {\oiii}$\lambda$5007. The black dashed hyperbola shows the edge of photo-ionized bi-cone, while the grey dashed hyperbola shows the edge of kinematic bi-cone. d: The spatially resolved BPT diagram, with the same color definition as in the panel b. The black dashed hyperbola shows the edge of the photo-ionized bi-cone.}
 \label{fig5}
\end{figure*}

In Figure \ref{fig1}c, we find that there are bi-cone like redshifted regions (marked by the grey dashed hyperbola) in the {\oiii}$\lambda$5007 velocity field. In this section, we further confirm the bi-cone features from both {\oiii}$\lambda$5007 EQW and BPT \citep{1981PASP...93....5B} diagrams. Figure \ref{fig5}b shows the BPT diagram for the spaxels whose S/Ns in the {\hb}, {\oiii}$\lambda$5007, {\ha} and {\nii}$\lambda$6583 are larger than 2. \cite{1981PASP...93....5B} demonstrated that it is possible to distinguish AGNs from normal star-forming galaxies by considering the intensity ratios of two pairs of relatively strong emission lines. The spaxels located in the AGN region are color-coded by the perpendicular distance from the \cite{2001ApJ...556..121K} line (black solid line in Figure \ref{fig5}b). Alternatively, we can separate the shock from the AGN emission using the demarcation line defined by \cite{2010ApJ...711..818S}. Figure \ref{fig5}d shows the resolved BPT diagram for SDSS J171359.00+333625.5. The blue pixels represent the star forming region, the green pixels mark the composite region of contribution from both AGN/shock and star formation, the yellow pixels trace the shock signal, while the orange pixels represent the AGN region. It is clear from this resolved BPT diagram that the disk of this galaxy is star forming (blue). The bi-cone region dominated by the AGN radiation (orange) with weak contamination from the shocks (yellow) is located perpendicular to the disk. The interface between galactic disk and bi-cone falls in the composite region (green).

However, we keep in mind that {\oiii}$\lambda$5007 has totally different kinematics (both in velocity and velocity dispersion) from the other three BPT lines (H$\alpha$, {\nii} and {\sii}), which may indicate that {\oiii} originates from the different regions. In Figure \ref{fig5}c, we show the {\oiii}$\lambda$5007 EQW map. The bi-cone region (marked by the black dashed hyperbola) with higher {\oiii}$\lambda$5007 EQW is obvious in this panel, and this bi-cone region is larger than the kinematic bi-cone outflows (grey dashed hyperbola from Figure \ref{fig1}c). We plot the {\oiii}$\lambda$5007 EQW map over the SDSS $g,r,i$ color image to give an idea about the scale of bi-cone, see Figure \ref{fig5}a. The estimated scale of bi-cone region is about 8.26 kpc, while the effective radius of SDSS J171359.00+333625.5 is only 3.99 kpc. This means that the bi-cone spans far beyond the galaxy disk, which also evidences that the contribution of star formation in ionization of bi-cone is negligible.

Combining the strong {\oiii}$\lambda$5007 emission region in Figure \ref{fig5}c and the AGN region (orange) in Figure \ref{fig5}d (see the black hyperbolas as guide lines), we conclude that the gas in the bi-cone outflows is primarily ionized by the central active black hole with negligible contribution from star formation or shocks. Taking {\oiii}$\lambda$5007 as a probe of NLR, we can estimate that the opening angle of cones is close to 80$^{\circ}$. Figure \ref{fig5}c also shows that the two cones in this galaxy are not symmetric with respect to the galactic mid-plane: the {\oiii} EQW in the eastern cone is stronger. Considering that this is not a totally edge-on galaxy with an inclination angle of 78$^{\circ}$, we tend to suggest that the eastern cone is in the front side of the disk. In this case, we observe the radiation from the eastern cone directly, while the strength of emission from the western cone is weakened after travelling through the clumps of gas and dust in the galaxy disk. Difference in the strength between {\oiii}$\lambda$5007 and the permitted emission lines, such as {\hb} and {\ha}, is primarily contributed by the disk. 
The extinction of the disk can be calculated using the flux ratio {\ha}$/${\hb}. 
It comes out that the {\ha}$/${\hb} ratio in the eastern part of the disk is
higher than in the west, which supports the assumption that the front cone
is in the east.

\section{Discussion}
\label{sec:discussion}
\begin{figure*}
 \resizebox{0.8\textwidth}{!}{\includegraphics{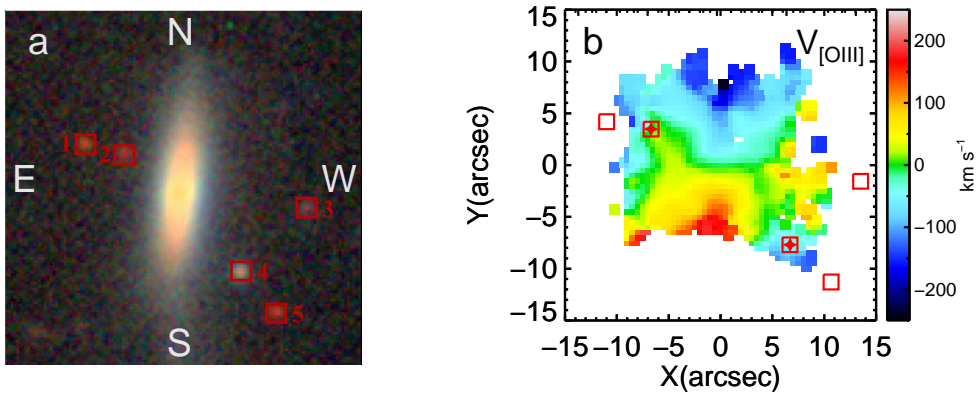}}
 \caption{a: A deep image of SDSS J171359.00+333625.5 obtained by 2.3-m Bok Telescope. The red squares mark five blobs around the galaxy. Two blobs labelled as $'$2 4$'$ are located within MaNGA bundle and share similar redshift with SDSS J171359.00+333625.5. b: The {\oiii}$\lambda$5007 velocity field for spaxels with {\oiii}$\lambda$5007 S/N larger than 3. The red squares mark the same positions as in the panel a. The red stars correspond to blobs $'$2 4$'$ which located within the {\oiii}$\lambda$5007 velocity field.}
 \label{fig6}
\end{figure*}

Figure \ref{fig6}a shows a deep (about two magnitudes deeper than SDSS in $g$-band observation) image of SDSS J171359.00+333625.5 obtained by 2.3-m Bok Telescope. From this panel, we find five blobs (marked by red squares) around our target. Two of these blobs are located within the MaNGA bundle (labelled by $'$2 4$'$), and the spectra of them show that the redshifts of these blobs are similar to SDSS J171359.00+333625.5. Figure \ref{fig6}b shows the {\oiii}$\lambda$5007 velocity field and we can find that some blueshifted {\oiii}$\lambda$5007 features overlap with two blobs (marked by red stars). There is also a blueshifted region without any obvious counterpart blob outside the eastern cone with X-position smaller than $-$7 arcsec. The {\oiii}$\lambda$5007 emission along the interface of this blueshifted region and redshifted eastern cone show double-peaked structures.

Galaxies with gas and stars counter-rotating are the key manifestations 
of the regulation by external processes, i.e. major mergers, minor mergers or gas accretion, which could bring counter-rotating gas into the galaxies. The existence of a large number of blobs around SDSS J171359.00+333625.5 should be the result of interaction, which also leads to the complicate kinematics of this galaxy. SDSS J171359.00+333625.5 holds abundant ionized gas which is accompanied by the recycle of feeding and feedback, thus it is a great laboratory for us to understand these physical processes in detail. However, with MaNGA data alone, we do not plan to over-interpret the observational results. Futher observations, like higher S/N and spatial resolved IFU data from VLT/MUSE and cold gas detection from ALMA, are required to uncover the mysteries of this galaxy. In this section, we summarise two most unusual properties of this galaxy, and discuss the possible explanations.

\subsection{Why is the Gas in the Bi-cone Region Redshifted?}

It is out of our expectation that the gas (tracted by {\oiii}$\lambda$5007, marked by grey dashed hyperbola in Figure \ref{fig1}c) in both the eastern and western cones are redshifted. In the typical AGN-driven winds picture, we would find a blueshifted cone in which the gas is moving towards the Earth and a redshifted cone where the gas is moving away. The two redshifted cones are also shown in Figure \ref{fig1}f (marked by black dashed hyperbola) where the disk model is subtracted from the {\oiii}$\lambda$5007 velocity field. In order to understand the origin of the redshifted kinematics in the two cones, we go through the {\oiii}$\lambda$5007 structures spaxel by spaxel, finding the evidence that the {\oiii}$\lambda$5007 emission lines have double peak structures in 10 $\sim$ 20 spaxels between the grey and black dashed hyperbolas in Figure \ref{fig1}c. However, both the resolution and S/N of these spectra are not high enough for us to separate the two kinematic components robustly. Thus, we can only list the possible explanations in this section.

\begin{figure*}
 \resizebox{0.43\textwidth}{!}{\includegraphics{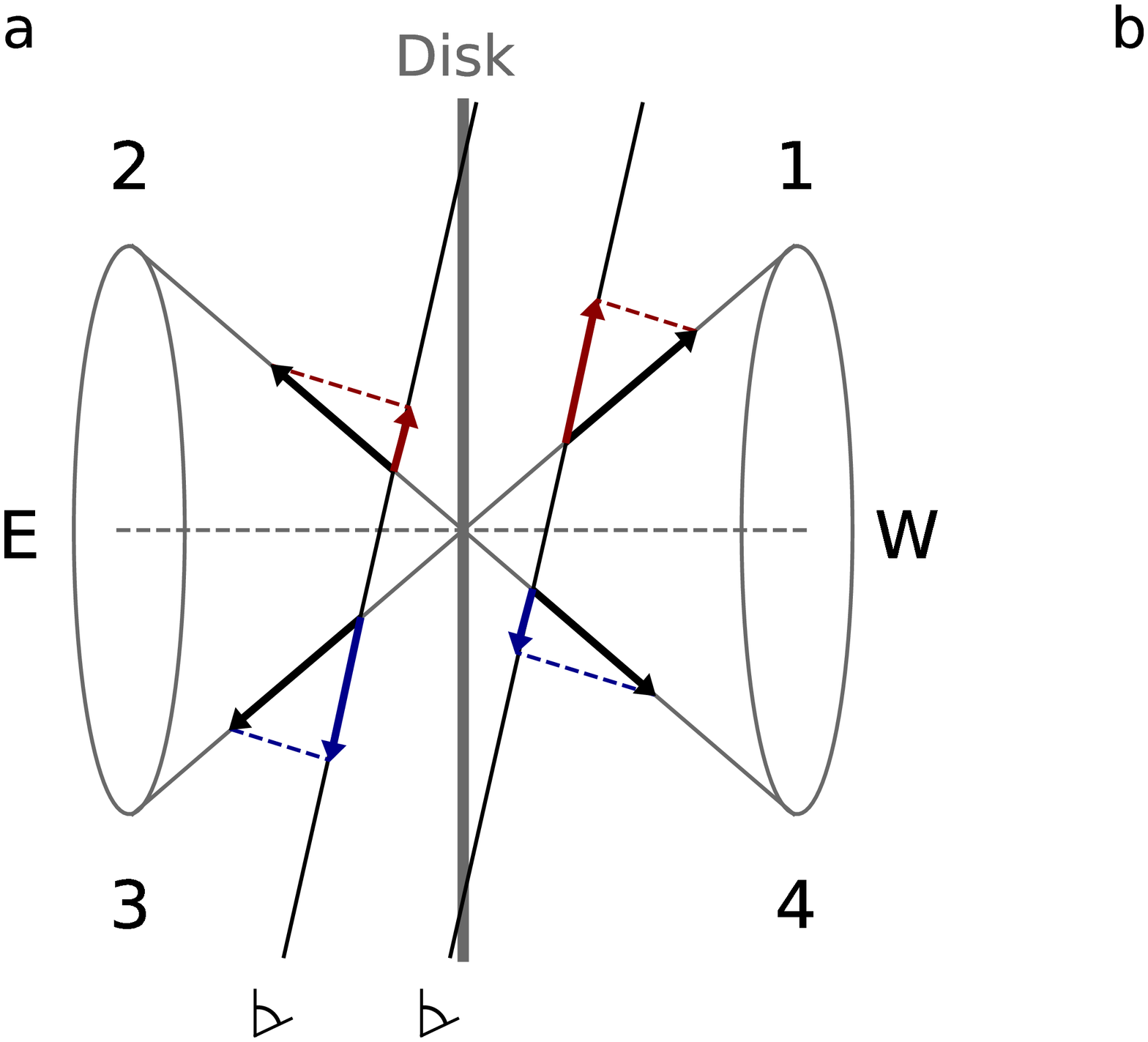}}
 \resizebox{0.135\textwidth}{!}{\includegraphics{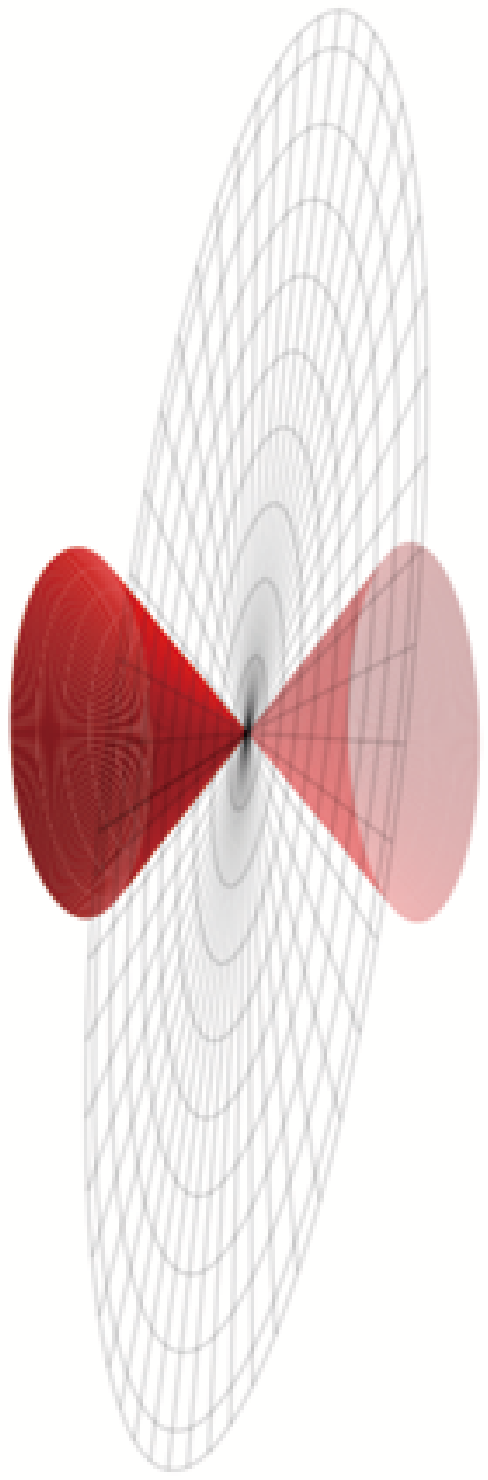}}
 \resizebox{0.415\textwidth}{!}{\includegraphics{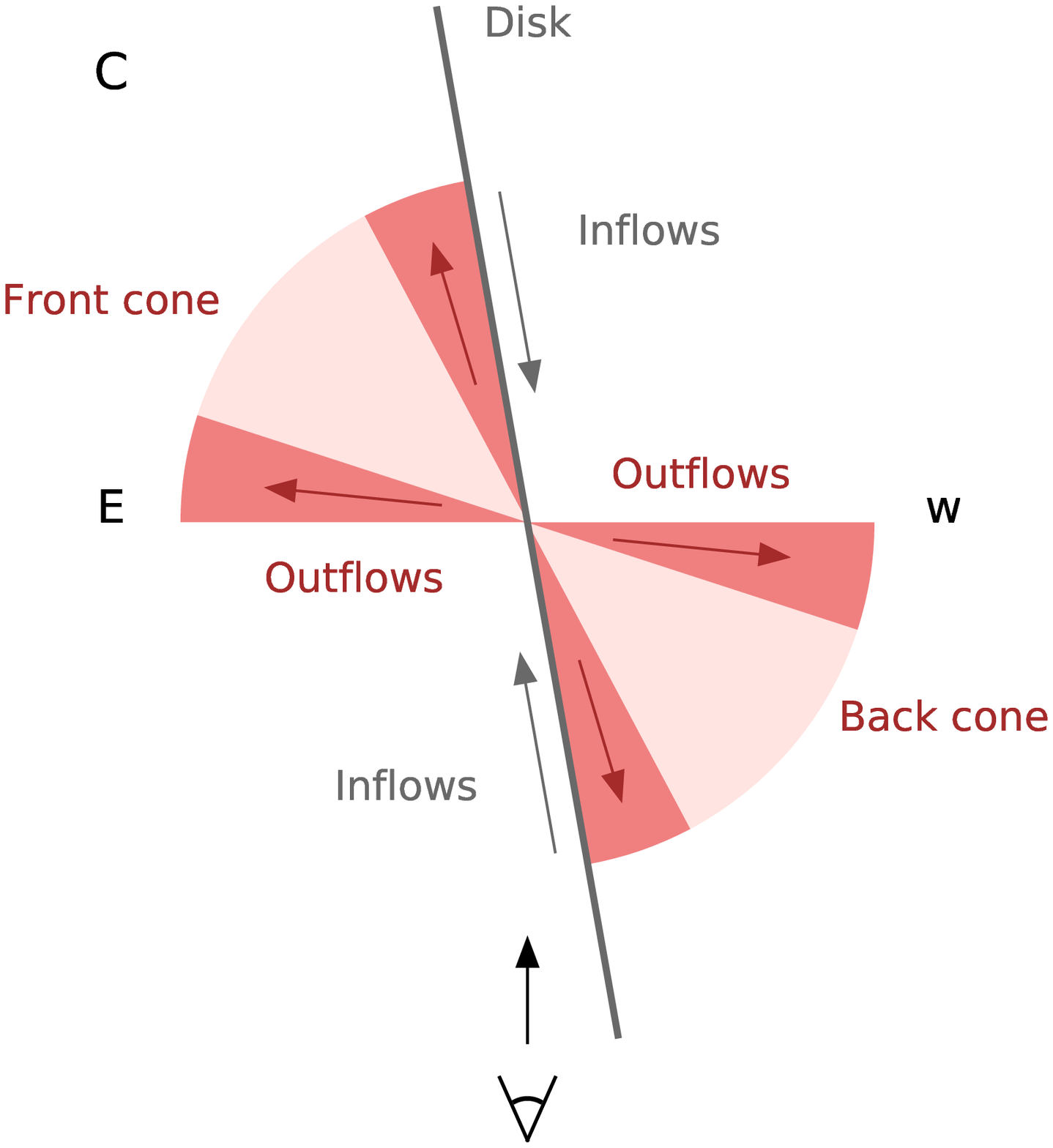}}
 \caption{a: Schematic drawing of winds along a bi-conical surface centered at the nucleus of a disk-shaped galaxy. The dashed horizontal line represents the cone’s symmetry axis along the minor-axis of the disk. The red arrows indicate the direction of redshifted velocity along the line-of-sight. The blue arrows show the blueshifted components. b: Geometric model of outflows and inflows, where the ellipse represents the disk, dark-red cone shows the front cone in the east, and the light-red cone shows the back cone in the west. c: Planform of panel b, the pink sectors represent the hollow regions close to the axis of the bi-cone.}
 \label{fig7}
\end{figure*}

One possible explanation is that the projected velocity field of gas is a combination of both galactic winds in the bi-cone and rotation of the disk. Figure \ref{fig7}a shows a schematic drawing of winds along the bi-conical surface centered on the nucleus of a disk-shaped galaxy \citep{1990ApJS...74..833H}, whose spatial structure is similar to M82 \citep{1988Natur.334...43B}. The cone's axis (dashed horizontal line) extends along the minor-axis of the disk. In the first quadrant (the top half of western cone), the projected velocity of winds along the line-of-sight is redshifted, and the disk rotation will lead to a blueshifted velocity along the line-of-sight. We find redshifted kinematics because this region is winds dominated, while the blueshifted disk hides behind the cone. The redshifted cone in the second quadrant (top-east) can be explained in a similar way. In the third and fourth quadrants (the bottom-east and bottom-west cones), the projected velocity is a combination of redshifted disk rotation and blueshifted winds, and it is disk dominated in these two quadrants since the disk is not blocked by the winds.

\cite{2015ApJ...799...83C} studied the outflows in NGC 4151 (quite similar to SDSS J171359.00+333625.5 in the spatial structure of bi-cone outflows and disk), and concluded that the peak mass outflow rate in NLR is 3.0 M$_{\sun}$ yr$^{-1}$. As we know, the gas inflow rate should be much larger than this value since most part of the gas has formed (or is forming) stars in the disk, another part has flown into the central black hole. Thus, we expect strong gas inflows in SDSS J171359.00+333625.5, which is just like the NGC 4151. The counter-rotating gas and stellar kinematics also supports this expectation since the collision between accreted counter-rotating gas and pre-existing gas will lead to the re-distribution of angular momentum and trigger strong gas inflows \citep{2016NatCo...713269C, 2016MNRAS.463..913J}. Figure \ref{fig6}a gives the evidence that SDSS J171359.00+333625.5 is a complicated system with not only galactic winds and disk components, but also the surrounding blobs that could be the source of accretion gas.

Considering the existence of strong inflows in this galaxy,  here comes another possibility that great amount of gas in the disk is being efficiently accreted into the central black hole, and the central AGN feedback is blowing the ionized gas out along the bi-cone. A geometric model and its planform are given in Figure \ref{fig7}b and \ref{fig7}c. An important point in this model is confirming the reality that the AGN-driven outflows are not necessarily  perpendicular to the disk (which has been illustrated in the latter part of introduction). Generally speaking, {\oiii}$\lambda$5007 in the east is mainly contributed by the front cone (dark-red cone in Figure \ref{fig7}b), and all the winds in this region are redshifted. The blueshifted inflows along the disk are too weak to affect the line center of {\oiii}$\lambda$5007, but can result in the double peaked structures in emission lines. The emission in the west is mainly contributed by the inflowing gas along the disk, which is close to the line-of-sight (lower grey arrow in Figure \ref{fig7}c). Another question we need to figure out is why all the blueshited winds seem to have negligible effect on the projected velocity field in the west. The $V$-band extinction reveals that disk is optically thick in this region, so the radiation from the western cone is greatly weakened.

\subsection{Is the Special Kinematics of Oxygen Element a Mystery?}

As discussed in \ref{Different [OIII] and Ha kinematics}, the behavior of {\oiii}$\lambda$5007 is really wired in both velocity and velocity dispersion. Figure \ref{fig1}c and \ref{fig1}d show that {\oiii}$\lambda$5007 and {\ha} share the same direction in rotation, but {\ha} rotates faster than {\oiii}$\lambda$5007. As we know, {\oiii}$\lambda$5007 is a forbidden line which is collision excited and can only exist in the low density region, while {\ha} emission is photo-ionization excited. If this is the origin of special kinematics of oxygen element, we would expect to find similar behavior in the other forbidden lines, i.e. {\nii}$\lambda\lambda$6548,6584 and {\sii}$\lambda\lambda$6716,6731. However, the truth is that only {\oii}$\lambda\lambda$3726,3729 and {\oiii}$\lambda$4959 share similar velocity and velocity dispersion with {\oiii}$\lambda$5007. While {\hb}, {\nii}$\lambda\lambda$6548,6584 and {\sii}$\lambda\lambda$6716,6731 have similar velocity and velocity dispersion to {\ha}. The second possibility is that {\oiii}$\lambda$5007 is more sensitive to shock than {\ha}. However, we keep in mind that we are presenting the velocity and velocity dispersion along major-axis where we do not expect strong shock generated from the interaction between winds and the interstellar medium. We have to admit that we do not quite understand the origin of the wired kinematics behavior of {\oiii}$\lambda$5007.

\section{Summary}
\label{sec:summary}

In this paper, we study the kinematic properties of a local edge-on Seyfert galaxy, SDSS J171359.00+333625.5, with X-shaped bi-cone ionized by AGN.
The spatially resolved data from MaNGA survey provide us an opportunity to uncover the specific kinematics of feeding and feedback. 

$\bullet$ The gas and stellar components are counter-rotating with respect to each other. Collision between the pre-existing and accreted gas can re-distribute the angular momentum and lead to
the gas inflow.

$\bullet$ The kinematics of oxygen traced by the 
{\oii}$\lambda$3727,3729 and {\oiii}$\lambda$4959,5007 lines, 
including the gas rotation velocity and velocity dispersion, 
is totally different from the other elements like hydrogen 
(as traced by {\hb} \& {\ha}), nitrogen (as traced by {\nii}$\lambda\lambda$6548,6584) and sulfur (as traced by {\sii}$\lambda\lambda$6716,6731). We notice that oxygen rotates slower than the other elements. On the other hand, nitrogen, sulfur and hydrogen follow the galaxy potential with higher velocity dispersion in the central region of bulge and decrease as radius increases towards the disk region, whereas the velocity dispersion of oxygen keeps being roughly a constant over the major-axis.

$\bullet$ The distribution of {\oiii} EQW is completely consistent with the X-shaped bi-cone, suggesting that the gas is primarily ionized by the central active black hole. The ionized gas in bi-cone region is receding with respect to the center.

$\bullet$ Two possible models can be used to explain the redshift in the bi-cone region. One is that the projected velocity field of gas is a combination of both winds in the bi-cone and the rotation of the disk. Another possibility is that great amount of gas in the disk is being efficiently accreted into the central black hole, and the blue-shifted winds in the west are totally obscured by the disk.\\

{\noindent \bf Acknowledgements}

This work is supported by the Y. C acknowledges support from the National Key R\&D Program of China (No. 2017YFA0402700), the National Natural Science Foundation of China (NSFC grants 11573013, 11733002, 11922302).
National Natural Science Foundation of China (Nos. 11873032, 11433005, 11173016) and by the Research Fund for the Doctoral Program of Higher Education of China (No. 20133207110006).
DB is partly supported by grant RSCF 19-12-00145.

Funding for the Sloan Digital Sky Survey IV has been provided by the Alfred P. Sloan Foundation, the U.S. Department of Energy Office of Science, and the Participating Institutions. SDSS- IV acknowledges support and resources from the Center for High-Performance Computing at the University of Utah. The SDSS web site is www.sdss.org. SDSS-IV is managed by the Astrophysical Research Consortium for the Participating Institutions of the SDSS Collaboration including the
Brazilian Participation Group, the Carnegie Institution for Science, Carnegie Mellon University, the Chilean Participation Group, the
French Participation Group, Harvard-Smithsonian Center for Astrophysics, Instituto de Astrof\'{i}sica de Canarias, The Johns Hopkins University, Kavli Institute for the Physics and Mathematics of the Universe (IPMU) / University of Tokyo, Lawrence Berkeley National Laboratory, Leibniz Institut  f\"{u}r Astrophysik Potsdam (AIP), Max-Planck-Institut  f\"{u}r   Astronomie  (MPIA  Heidelberg), Max-Planck-Institut   f\"{u}r   Astrophysik  (MPA   Garching),
Max-Planck-Institut f\"{u}r Extraterrestrische Physik (MPE), National Astronomical Observatory of China, New Mexico State University, New York University, University of Notre Dame, Observat\'{o}rio Nacional / MCTI, The Ohio State University, Pennsylvania State University, Shanghai Astronomical Observatory, United Kingdom Participation Group, Universidad Nacional  Aut\'{o}noma de M\'{e}xico,  University of Arizona, University of Colorado  Boulder, University of Oxford, University of Portsmouth, University of Utah, University of Virginia, University  of Washington,  University of  Wisconsin, Vanderbilt University, and Yale University.

\bsp

\label{lastpage}
\end{document}